\documentclass[twocolumn,showpacs,preprintnumbers,amsmath,amssymb,prl]{revtex4-1}
\usepackage{graphicx}
\usepackage{color}
\usepackage{amsmath,amsthm,amssymb}
\usepackage{braket}
\begin{document}
 
\title{Antiferromagnetic Ground State of La$_{2}$CuO$_{4}$: A Parameter-free Ab Initio Description}

\author{Christopher Lane*\textsuperscript{1}}
\author{James W. Furness\textsuperscript{2}}
\author{Ioana Gianina Buda\textsuperscript{1}}
\author{Yubo Zhang\textsuperscript{2}}
\author{Robert S. Markiewicz\textsuperscript{1}}
\author{Bernardo Barbiellini\textsuperscript{3,1}}
\author{Jianwei Sun*\textsuperscript{2}}
\author{Arun Bansil*\textsuperscript{1}}

\affiliation{
\textsuperscript{1}Physics Department, Northeastern University, Boston MA 02115, USA \\
\textsuperscript{2}Department of Physics and Engineering Physics, Tulane University, New Orleans, LA 70118, USA\\
\textsuperscript{3}Department of Physics, School of Engineering Science, Lappeenranta University of Technology, FI-53851 Lappeenranta, Finland\\
}

\date{version of \today} 
\begin{abstract}
We show how an accurate first-principles treatment of the antiferromagnetic (AFM) ground state of La$_2$CuO$_4$ can be obtained without invoking any free parameters such as the Hubbard $U$. The magnitude and orientation of our theoretically predicted magnetic moment of $0.495 \mu_{B}$ on Cu-sites along the (100) direction are in excellent accord with experimental results. The computed values of the band gap (1.00 eV) and the exchange-coupling (-138 meV) match the corresponding experimental values. We identify interesting band splittings below the Fermi energy, including an appreciable Hund's splitting of 1.25 eV. The magnetic form factor obtained from neutron scattering experiments is also well described by our calculations. Our study opens up a new pathway for first-principles investigations of electronic and atomic structures and phase diagrams of cuprates and other complex materials. 
\end{abstract}

\pacs{}

\maketitle 
A fundamental challenge that has remained unsolved ever since the discovery of high-$T_c$ superconductivity in hole-doped La$_2$CuO$_4$ nearly 30 years ago has been that a first-principles description of the ground state electronic structure of La$_2$CuO$_4$ has not been possible. The magnetic state of La$_2$CuO$_4$, in particular, has been especially hard to capture within a uniform theoretical picture. Various attempts within the Hohenberg-Kohn-Sham\cite{hohenberg_inhomogeneous_1964,kohn_self-consistent_1965} density functional theory (DFT) framework have at best yielded mixed results\cite{pickett_electronic_1989}. In particular, most studies have struggled to model correctly the antiferromagnetic (AFM) ground state of La$_2$CuO$_4$, and have therefore been unable to provide a handle on the key experimentally observed properties of this parent compound, which gives birth to the novel phenomena of high-T$_{c}$ superconductivity. 

More specifically, the local-spin-density-approximation\cite{jones_density_1989,perdew_self-interaction_1981} incorrectly predicts La$_2$CuO$_4$ and other half-filled cuprates to be  nonmagnetic (NM)  metals\cite{yu_electronically_1987,mattheiss_electronic_1987,pickett_electronic_1989,ambrosch-draxl_local-spin-density_1991} in complete disagreement with experimental findings. The generalized gradient approximation (GGA)\cite{perdew_generalized_1996} only produces a weak AFM order\cite{singh_gradient-corrected_1991}. While Hartree-Fock captures the AFM ground state and the magnetic form factor, the computed band gap of $\approx$17 eV is far too large\cite{uchida1991optical,ono_strong_2007}, and the strength of the exchange coupling is too small by a factor of four\cite{su_crystal_1999}. 

Failure of the DFT in capturing the AFM state of half-filled cuprates has led to the widely held belief that DFT is fundamentally limited in its reach for addressing electronic structures of cuprates and many other classes of important materials. The development of methods, which incorporate stronger electron correlations in order to stabilize the AFM ground state, has been effective in describing the low-energy spectra of the cuprates. These include `beyond DFT' schemes for extending the DFT into the intermediate coupling regime\cite{das_intermediate_2014} such as the quasi-particle GW (QP-GW) and various dynamical mean field theory (DMFT) based schemes \cite{kotliar_electronic_2006,held_realistic_2006,park_cluster_2008}. All beyond DFT schemes, however, require the introduction of empirically derived, ad hoc parameters, which compromise their predictive power.

{Following the theorems of Hohenberg-Kohn and Kohn-Sham, there must exist an exact exchange-correlation energy ($E_{xc}$) functional that incorporates {\it all} many-body effects into an effective single-particle Hamiltonian\cite{hohenberg_inhomogeneous_1964,kohn_self-consistent_1965,kohn1999nobel}. This would allow an exact {\it ab initio} treatment of all materials, including strongly correlated systems, at least insofar as the ground state energy and the related physical properties are concerned. As we invoke improved approximations to the exchange-correlation functional, we can then also expect concomitant improvements in the DFT predictions of the ground state properties\footnote{{See Section I of the Supplementary Materials for further discussion}}

Recently, the strongly-constrained-and-appropriately-normed (SCAN) meta-GGA exchange-correlation functional\cite{sun_strongly_2015}, which obeys all known constraints applicable to a meta-GGA functional\footnote{ {For technical implementation aspects of meta-GGA functionals, see Sec. II of the Supplementary Materials.} }, has shown promise\cite{car2016density} by significantly improving the description of diverse systems. These include surface properties of metals\cite{patra2017properties}, ice\cite{sun2016accurate} and liquid water\cite{chen2017ab}, subtle structural distortions in ferroelectrics\cite{paul2017accuracy,zhang2017comparative}, and transitions from insulator and semiconducting to metalic phases\cite{sun2016accurate,furness_accurate_nodate}.

}

\begin{figure*}[ht!]
\includegraphics[scale=0.31]{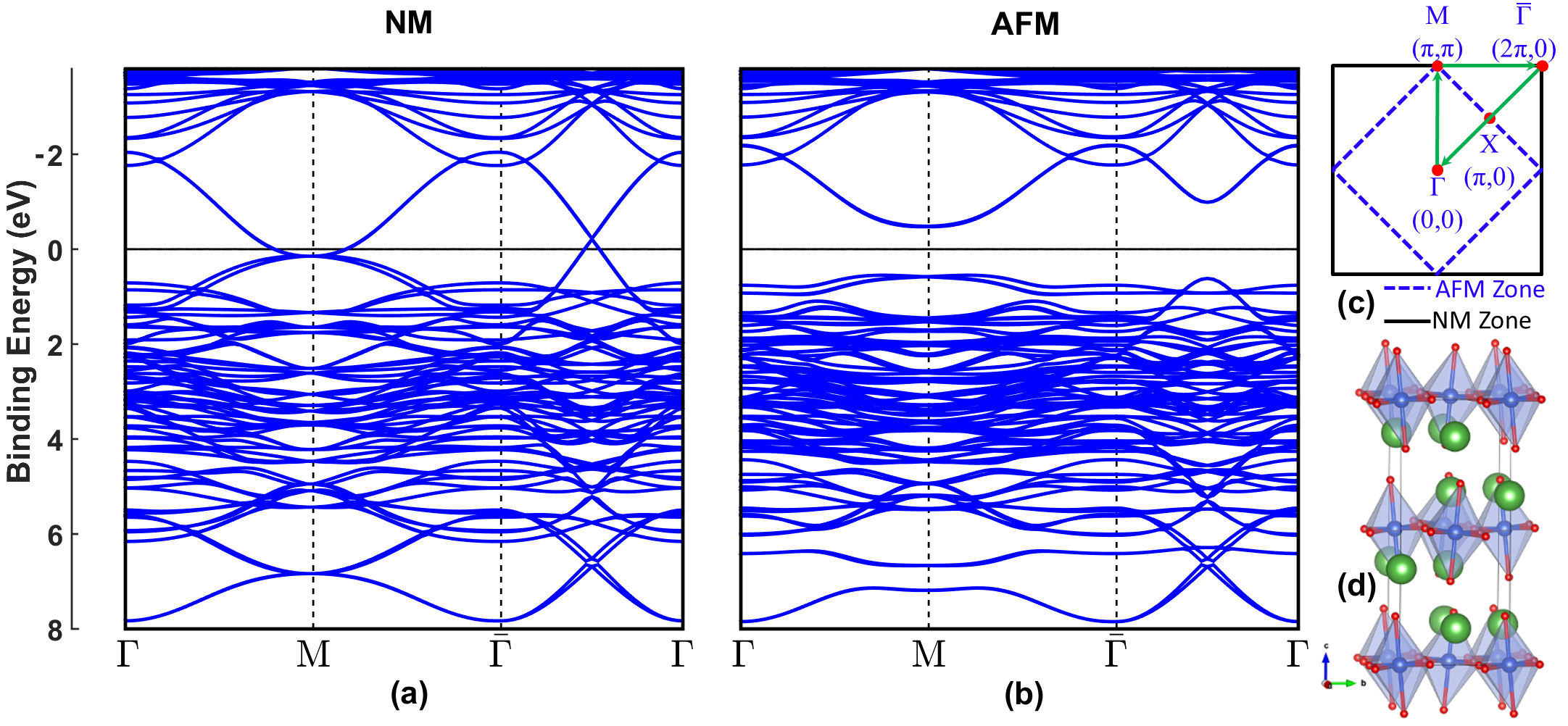}
\caption{(color online) (a,b) Electronic band dispersions of  La$_{2}$CuO$_{4}$  in the LTO crystal structure for the nonmagnetic (NM) and antiferromagnetic (AFM) phases. (c) A schematic of the NM and AFM Brillouin zones with the path followed by the electronic dispersions in panels (a) and (b) highlighted. (d) Crystal structure of La$_2$CuO$_4$ in the LTO phase with copper, oxygen and lanthanum represented by blue, red and green spheres respectively. }
\label{fig:BANDS}
\end{figure*}

In this Letter, we show how electronic and magnetic structures of La$_2$CuO$_4$ can be captured accurately by SCAN meta-GGA\cite{sun_strongly_2015} within the DFT framework. Our first-principles, parameter-free magnetic ground state obtained in this way reproduces the key experimentally observed properties of La$_2$CuO$_4$. These include the magnitude and {orientation} of the local magnetic moment on copper-sites, size of the optical band gap, strength of the exchange-coupling and the shape of the magnetic form factor. {The accuracy of these predictions reflects the systematic improvement in the exchange-correlation energy embodied in the SCAN functional. }

{\it Ab initio} calculations were carried out by using the pseudopotential projector-augmented wave method\cite{kresse_ultrasoft_1999} implemented in the Vienna ab initio simulation package (VASP) \cite{kresse_efficient_1996,kresse_ab_1993} with an energy cutoff of 500 eV for the plane-wave basis set. Exchange-correlation effects were treated using the SCAN meta-GGA scheme\cite{sun_strongly_2015}, where a 12 x 12 x 6 $\Gamma$-centered k-point mesh was used to sample the Brillouin zone. Spin-orbit coupling effects were included self-consistently. We used the low-temperature-orthorhombic (LTO) crystal structure of $Bmab$ symmetry in accord with the experimentally observed structure of La$_2$CuO$_4$. \cite{jorgensen_superconducting_1988, ginsberg1998physical,furness_accurate_nodate} All sites in the unit cell along with the unit cell dimensions were relaxed using a conjugate gradient algorithm to minimize energy with an atomic force tolerance of $0.008$ eV/\AA~and a total energy tolerance of $10^{-5}$ eV. The theoretically obtained structural parameters are in good accord with the corresponding experimental results, see Sec. VI of the Supplementary Material for details. As shown in Fig. \ref{fig:BANDS}(d), the LTO structure can be viewed as being a $\sqrt{2} \times  \sqrt{2}$ body-centered-tetragonal superlattice of I4/mmm symmetry in which $a^{\prime}\approx b^{\prime}\approx \sqrt{2}a$; the CuO$_{6}$ octahedra are rotated along the (110) and (1$\bar{\text{1}}$0) directions in alternate layers.

\begin{figure*}[ht]
\includegraphics[scale=0.305]{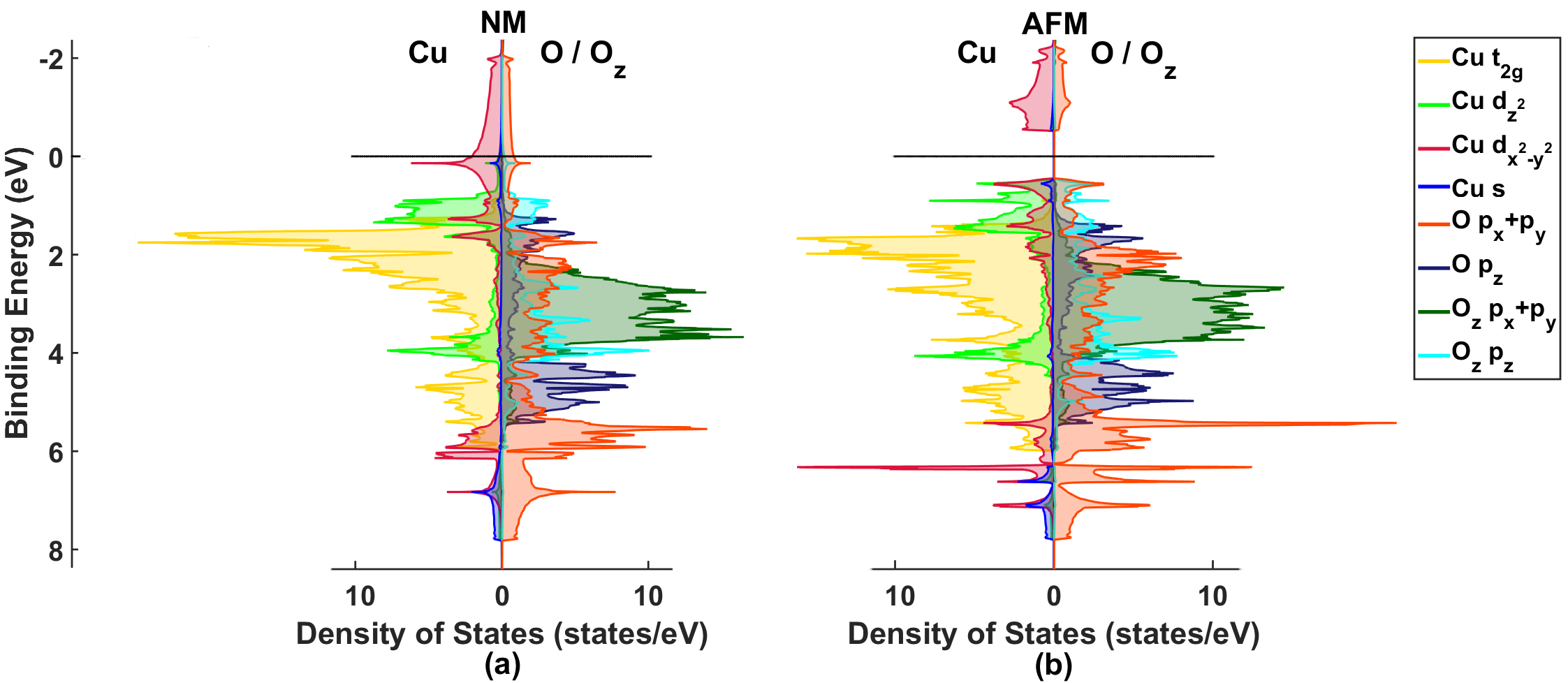}
\caption{(color online) Site-resolved partial densities of states in the nonmagnetic (NM) and antiferromagnetic (AFM) states of La$_{2}$CuO$_{4}$ in the LTO structure. Copper and oxygen characters are plotted on the left and right hand sides, respectively. Shadings and lines of various colors (see legend) give contributions from various orbitals of copper, apical (O$_{z}$) and in-plane (O) oxygen sites.}
\label{fig:DOS}
\end{figure*}


Figures \ref{fig:BANDS}(a) and \ref{fig:DOS}(a) show the band structure and partial densities-of-states (PDOSs) associated with various Cu and O orbitals in the NM phase{\footnote{ {{The energy of the NM state is $0.593$ eV/unit cell higher than that of the AFM state, and it is for this reason not relevant for low-energy physics. It is shown only for the purpose of comparison.}} }}. Here, and throughout, we will distinguish the in-plane oxygen atoms from the apical oxygen atoms as O and O$_{z}$, respectively. We see that the half-filled anti-bonding Cu d$_{x^{2}-y^{2}}$/O p$_{x}$+p$_{y}$ band crosses the Fermi level and that its bonding partner `bookends' the PDOS from the bottom over binding energies of $6-8$ eV; see Sec. X in the Supplementary Material for orbital contributions to band structure. The orbital character at the Fermi level is predominantly ($70\%$) Cu d$_{x^{2}-y^{2}}$, with O p$_{x}$+p$_{y}$, Cu d$_{z^{2}}$, O$_{z}$ p$_{z}$ and Cu $s$ sharing the remaining ($30\%$) weight. On the other hand, at binding energy of $6.8$ eV, the character is mainly O p$_{x}$+p$_{y}$ ($52\%$), Cu d$_{x^{2}-y^{2}}$ (19\%) and Cu $4$s  (14\%). Preceding results are similar to what we would expect from a molecular-bonding type picture of an octahedrally coordinated metal. \cite{goodenough_magnetism_nodate,rao_transition_1998} Note that due to the tetragonal Jahn-Teller distortion of the Cu-O octahedra, the Cu d$_{z^{2}}$/O$_{z}$-p$_{z}$ anti-bonding level lies at approximately $1$ eV while the related bonding level lies around $4$ eV below the Fermi energy. Copper d$_{z^{2}}$ and apical oxygen p$_{z}$ have a combined weight of 70\% and 60\% of the total DOS at 1 eV and 4 eV, respectively. Remaining states in the crystal-field-split manifold, i.e. the non-bonded oxygen atoms and the hybridized t$_{2g}$ levels, sit at binding energies of $1-6$ eV. In comparison to the usual hybridization schematic by Fink et. al.\cite{fink_electronic_1989}, we highlight the non-negligible presence of Cu d$_{z^{2}}$ and 4$s$ in the molecular-bonding picture of copper and oxygen; see Sec. VII in the Supplementary Material for details.

Figures \ref{fig:BANDS}(b) and \ref{fig:DOS}(b) show that the AFM state stabilizes with a band gap of approximately $1$ eV that opens up around the Fermi energy of the NM system. This gap, in good agreement with optical\cite{uchida1991optical} {and transport\cite{ono_strong_2007} data (see Sec. IV of the Supplementary Materials for details)}, develops in the half-filled Cu d$_{x^{2}-y^{2}}$ dominated band by splitting the up and down spin anti-bonding bands. Remarkably, as a result of electron-electron interactions, a `mirrored' splitting occurs around $7$ eV in the bonding band, which breaks its spin degeneracy (see orbital contributions to the AFM band structure under Sec. X of the Supplementary Material). This splitting occurs along the $\Gamma-M-\bar{\Gamma}$ cut in the Brillouin zone forming a $0.5$ eV gap. However, due to the strong O p$_{x}$+p$_{y}$ character along $\bar{\Gamma}-\Gamma$, a full gap in the energy spectrum is prevented. A further consequence of this splitting is the generation of a flat (non-dispersing) band, exhibiting a strong van Hove singularity in the DOS. Distinct splittings are also generated in the Cu d$_{z^{2}}$/O$_{z}$ p$_{z}$ bonding and anti-bonding bands. As a result of splittings in the anti-bonding d$_{z^{2}}$ bands a gap of $0.16$ eV forms, seen at $1$ eV along $\Gamma-M$ in the electronic structure (see Sec. IX in the Supplementary Materials for details). There may be similar splittings within the $t_{2g}$ complex, but these are harder to discern. 

{ The aforementioned spin-splittings can be seen as being a consequence of intrasite multi-orbital electron-electron interactions. Through simple manipulations of the site-resolved partial density of states in combination with the multi-orbital Hubbard model\cite{oles1983antiferromagnetism} (see Sec. VIII of the Supplementary Material), we have estimated effective Hund's and intra-orbital potentials for various Cu 3d orbitals. In this way, we obtain $U$ to be $4.846$ eV and an orbital-dependent Hund's splitting ($J_{H}$) with an upper-bound of $1.248$ eV. These values are inline with Jang et al. [\onlinecite{jang2016direct}] and suggest that LCO is closer to a Slater type insulator in agreement with Comanac et al. [\onlinecite{comanac2008optical}] and SU(2) spin models\cite{paramekanti2007n}. }

In the AFM phase, the conduction states are dominated by d$_{x^{2}-y^{2}}$ (68\%). However, the valence states are not dominated by Cu d$_{x^{2}-y^{2}}$, but consist of almost equal contributions (around $0.55$ eV) from d$_{x^{2}-y^{2}}$/O p$_{x}$+p$_{y}$ and d$_{z^{2}}$/O$_{z}$ p$_{z}$ ($\approx$20\% each). The unexpected character of the valence states stems from an appreciable splitting in the d$_{z^{2}}$/O$_{z}$ p$_{z}$ anti-bonding level. We emphasize that, {due to the sizable $d_{z^2}$ contribution to the valence states,} the conventional one-band model of the cuprates is of limited reach\cite{sakurai2011imaging}, as is the classification of the cuprates within the Zaanen-Sawatzky-Allen\cite{zaanen_band_1985} scheme. 


\begin{figure}
\includegraphics[scale=.25]{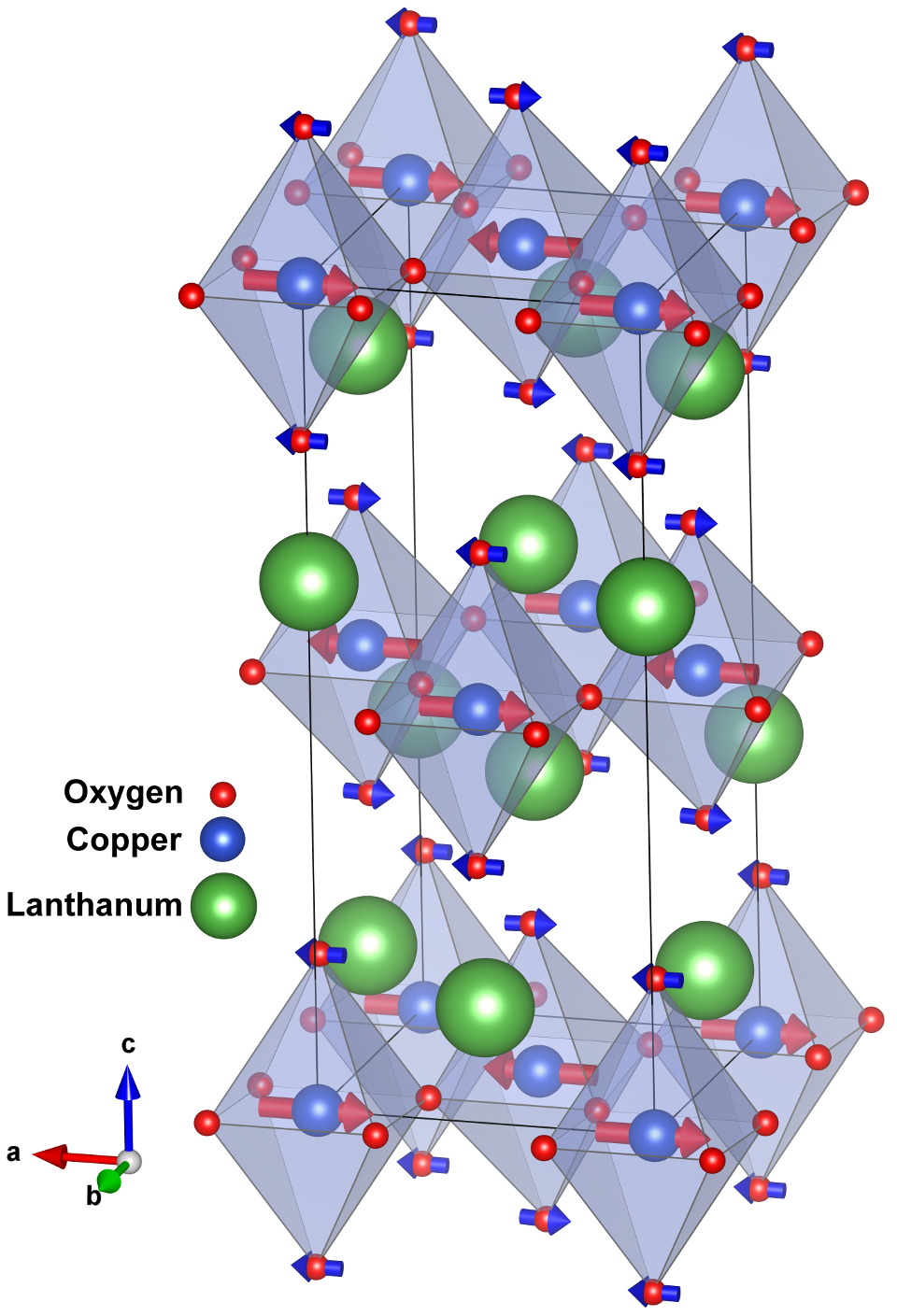}
\caption{(color online) Theoretically predicted AFM state of La$_2$CuO$_4$ in the LTO crystal structure. Red and blue arrows represent copper and apical oxygen magnetic moments, respectively; in-plane oxygen atoms have no net magnetic moment. Octahedral faces are shaded in blue; black lines mark the unit cell. }
\label{fig:SPINSTRUCTURE}
\end{figure}

Figure \ref{fig:SPINSTRUCTURE} shows our theoretically obtained AFM structure within the LTO unit cell, where red and blue arrows denote copper and apical oxygen magnetic moments, respectively. The predicted value of the magnetic moment on copper sites is $0.495 \mu_{B}$\footnote{The local magnetic moment of $0.495\mu_{B}$ is calculated by integrating the magnetic moment within a PAW sphere of radius $2.20$ \AA.}, which is in accord with the corresponding experimental value of {$0.60\pm0.05 \mu_{B}$\cite{tranquada2007neutron,yamada1987effect}(see Sec. V of the Supplementary Materials for more discussion of the variability in experimental values)}. Moreover, the copper magnetic moment vector in Fig. \ref{fig:SPINSTRUCTURE} clearly displays the planar { Ising} AFM ordering along the (100)-axis as seen in { low temperature} experimental studies\cite{kastner_magnetic_1998}. Our calculations show that the delicate $2^{\circ}$ out-of-the-plane spin tilt\cite{freltoft_magnetic_1988,kastner_magnetic_1998} is energetically indistinguishable  from the $0^{\circ}$ orientation. { The pinning of the moment vector to a fixed lattice direction would not be possible without the inclusion of spin-orbit coupling\footnote{ { Since the low temperature spins in LCO are experimentally seen to be Ising-like, we expect to reasonably capture fluctuation effects in our computed moments.} }} We obtain a small moment ($0.01 \mu_{B}$) on the apical oxygen (blue), which is anti-collinear to that of copper atoms lying at the centers of the octahedra. The in-plane oxygen atoms exhibit { spin polarization but} no net magnetic moment. { Here the performance of the SCAN functional better captures the subtle effects of many-body interactions and hybridizations in the solid state environment compared to previous semi-local functionals.\footnote{ { Our value of the moment is in good agreement with QMC calculations \cite{wagner_effect_2014}, although our model goes a step beyond these calculations by including the spin-orbit pinning of the moment to the lattice.} }   }  


In order to determine the strength of the exchange coupling, we map the total energies of the AFM and FM phases onto those of the nearest-neighbor-spin $\frac{1}{2}$ Heisenberg Hamiltonian in the mean-field approximation\cite{bedell_high_1990,noodleman_valence_1981,su_crystal_1999}. For La$_2$CuO$_4$, the Heisenberg Hamiltonian gives a reasonable description of the low-lying excitations, and thus a good estimate of the Heisenberg exchange parameter $J$\cite{kastner_magnetic_1998}. In the mean-field limit, the difference in total energies of the FM and AFM phases is
\begin{equation}\label{eq:heis_meanfield}
\Delta E= E_{AFM}-E_{FM}= JNZ\braket{S}^{2},
\end{equation}
where $N$ is the total number of magnetic moments, $S$ is the spin on each site, and $Z$ is the coordination number. The in-plane interactions within the Cu-O planes in La$_2$CuO$_4$ are much stronger than the inter-planar interactions, so that we can take $Z=4$. Since we normalize to one formula unit, $N=1$. Using the total energies for FM and AFM states obtained from our first-principles computations then yields $J=-138$ meV, where spin-orbit coupling is found to further stabilize the AFM state by $2.5$ meV\footnote{ { We expect a direct first-principles computation of the charge and magnetic susceptibilities of our AFM ground state of La$_2$CuO$_4$ will reproduce the spin-wave spectrum reasonably, as was demonstrated by Savrasov\cite{savrasov1998linear} and from tight-binding fits to LDA within many-body perturbation theory\cite{schrieffer1988spin,singh1990collective}. } } The present estimate of $J$ is in excellent accord with the experimentally determined $J$ value of $-133\pm 3$ meV\cite{bourges_superexchange_1997,coldea_spin_2001}, and represents a substantial improvement over previous Hartree-Fock calculations\cite{su_crystal_1999}.


{A test of the efficacy of our first-principles modeling is the reproduction of the experimental magnetic form factor\cite{pickett_electronic_1989},  since neutrons probe the local, microscopic magnetism in condensed matter systems. }The neutron magnetic cross-section can be factored into the dynamical spin-correlation function, $S(q,\omega)$, and the squared magnitude of the magnetic form factor, $|F(q)|^{2}$, where $F(q)$ probes effects of the magnetization cloud associated with each magnetic scattering center \cite{furrer_neutron_2009}. $F(q)$ in La$_2$CuO$_4$ has been assumed to resemble that of atomic Cu$^{2+}$, with deviations due to covalency being large enough to be observable \cite{walters_effect_2009,pickett_electronic_1989} { and to give a strong contribution to the exchange coupling\cite{bourges_superexchange_1997,coldea_spin_2001}}. We obtain the magnetic form factor from our spin-dependent charge densities via,
\begin{equation}\label{eq:heis_meanfield}
F(\mathbf{q})=\int d^{3}r~e^{i\mathbf{k}\cdot \mathbf{r}}\rho_{s}(\mathbf{r}), 
\end{equation}
by taking the Fourier transform of the spin density, $\rho_{s}(\mathbf{r})$, which is given by $\rho_{\uparrow}(\mathbf{r})-\rho_{\downarrow}(\mathbf{r})$, or the difference of the up and down spin densities. 

\begin{figure}
\includegraphics[scale=.32]{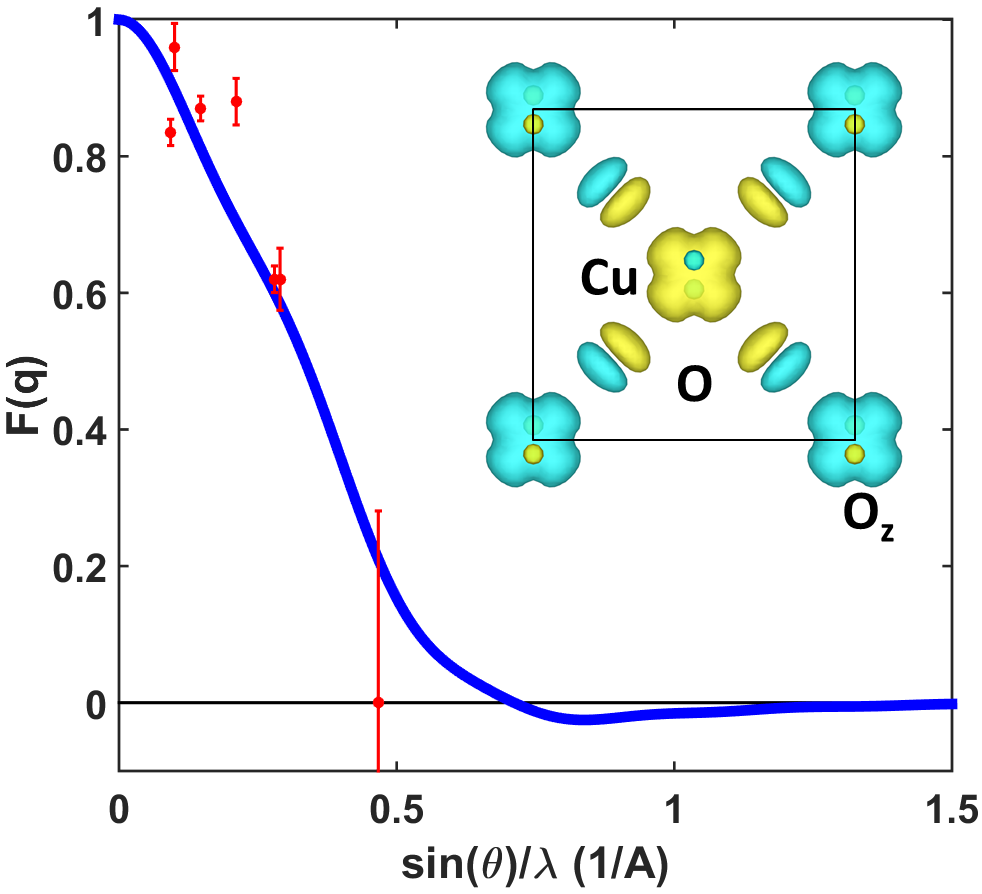}
\caption{(color online) Theoretical (blue line) and experimental (red dots with error bars, after Ref. \onlinecite{freltoft_magnetic_1988}) magnetic form factors for the AFM ground state of La$_2$CuO$_4$ in the LTO crystal structure. (Inset) Spin-density isosurface of the Cu-O plane. Yellow (blue) colors denote positive (negative) spin density; black lines mark the unit cell.}
\label{fig:FORMFACTOR}
\end{figure}

Figure \ref{fig:FORMFACTOR} compares the calculated form factor (blue symbols) with the available experimental values\cite{freltoft_magnetic_1988} (red points). The theoretical $F(q)$, which includes hybridization effects, is seen to be in reasonable accord with the experimental {line shape, implying a hybridization strength in accord with experiment}. The lineshape is similar to that of Walters {\it et al.} [\onlinecite{walters_effect_2009}], who account for hybridization explicitly in terms of model Wannier functions{\footnote{{Notably, we do not {\it a priori} assume the orbital character for the density used in calculating the magnetic form factor as was done by Walters {\it et al.} [\onlinecite{walters_effect_2009}]. By using our self-consistent magnetic density, we automatically include contributions from orbitals beyond Cu $d_{x^2-y^2}$ and O $p_x$ , $p_y$}}}. Our results yield a significant improvement over previous magnetically constrained studies\cite{leung_band-theoretical_1988} in which the form factor differed from the atomic and experimental lineshapes and predicted a peak between $0.1-0.2$ \AA$^{-1}$.

Figure \ref{fig:FORMFACTOR}(inset) shows the spin-density isosurface of the Cu-O plane, where yellow (blue) colors denote positive (negative) spin density. The magnetic moment is centered on the copper sites, with the polarization alternating in a checkerboard antiferromagnetic pattern. The magnetic moment is seen to spread from the copper atoms onto the in-plane oxygen atoms through hybridization effects. As a result, the in-plane oxygen atoms { develop a spin polarization, wherein the spin up orbital has, e.g., {\it $s+p_x$} symmetry and the spin down has {\it $s-p_x$} symmetry,} with zero net moment, and the magnetization in the Cu-O plane develops a quadrupole form. {Such an effect requires partly filled O orbitals, and hence considerable Cu-O hybridization.}  The opposing moment of the apical oxygen atoms is seen above the centers of copper sites.

Further insight is obtained by breaking down the magnetization at various sites into orbital contributions. On copper sites, the moment is dominated by the $d_{x^2-y^2}$ orbital with $-0.511 \mu_{B}$ with an opposing $s$ contribution of $+0.016 \mu_{B}$. No contributions from the $d_{z^2}$ or the charge saturated $t_{2g}$ manifold were found. The moment on the apical oxygen has mainly a $p_{z}$ character, with a moment of $+0.008 \mu_{B}$.


In conclusion, we have demonstrated clearly that an accurate first-principles treatment of the magnetic structure of the AFM ground state of La$_2$CuO$_4$ as an exemplar half-filled high-temperature cuprate superconductor is possible without invoking any free parameters such as the Hubbard $U$. Our study opens a new pathway for examining electronic structures and phase diagrams of cuprates{\footnote{{Our preliminary results using SCAN on YBCO, Hg-compounds, and Tl-cuprates also indicate that the SCAN functional stabilizes the AFM ground state of the half-filled systems.}} } and other complex materials, including magnetic phases, and the evolution of electronic spectra with pressure and doping and the related phenomena.


This work was supported (testing efficacy of new functionals in complex materials) by the DOE Energy Frontier Research Centers: Center for the Computational Design of Functional Layered Materials (DE-SC0012575). The work at Northeastern University was also supported by the U.S. Department of Energy (DOE), Office of Science, Basic Energy Sciences (grant number DE-FG02-07ER46352) (core research) and benefited from Northeastern University's Advanced Scientific Computation Center, the National Energy Research Scientific Computing Center supercomputing center (DOE grant number DE-AC02-05CH11231). The work at Tulane University was also supported by the startup funding from Tulane University.

*Corresponding authors: Christopher Lane (c.lane@neu.edu), Jianwei Sun (jsun@tulane.edu), Arun Bansil (ar.bansil@neu.edu)

\bibliography{LCO_refs}

\end{document}